\documentclass[aps,prx,reprint,twocolumn,noshowpacs,floatfix,longbibliography]{revtex4-1}  % for review and submission
\usepackage[]{graphicx}
\usepackage[%
  colorlinks=true,
  urlcolor=blue,
  linkcolor=blue,
  citecolor=blue
]{hyperref}
\usepackage{floatrow}
\usepackage{amssymb}
\usepackage{amsmath}
\graphicspath{{../img/all/}}
\usepackage{soul}
\soulregister\cite7
\soulregister\ref7
\usepackage[dvipsnames]{xcolor}
\usepackage{empheq}
\DeclareMathOperator*{\argmin}{argmin} % thin space, limits underneath in displays
\DeclareMathOperator*{\argmax}{argmax} % thin space, limits underneath in displays
\newcommand\SB[1]{\textsubscript{#1}}
\definecolor{mycolor}{RGB}{255,226,173}

\begin{document}
\title{Long-time Evolution of Interfacial Structure of Partial Wetting}
\author{Mengfei He}
\affiliation{\textit{Department of Physics, the James Franck and Enrico Fermi Institutes, the University of Chicago, Illinois 60637, USA}}
\begin{abstract}
When a solid plate is withdrawn from a partially wetting liquid, a liquid layer dewets the moving substrate.  High-speed imaging reveals alternating thin and thick regions in the entrained layer in the transverse direction at steady state.  This paper systematically compares this situation to the reversed process, forced wetting, where a solid  entrains an air layer along its surface as it is pushed into a liquid.  To quantify the absolute thickness of these steady-state structures precisely, I have developed an optical technique, taking advantage of the angle dependence of interference, combined with a method based on a maximum likelihood estimation.  The data show that the thicknesses of both regions of the film scale with the capillary number, $\mathrm{Ca}$.  In addition, a new region is observed during onset which differs from the behavior predicted by previous models.
\end{abstract}
\maketitle

\section{Introduction}  
\begin{figure}[htp]
	\includegraphics[width=0.8\textwidth]{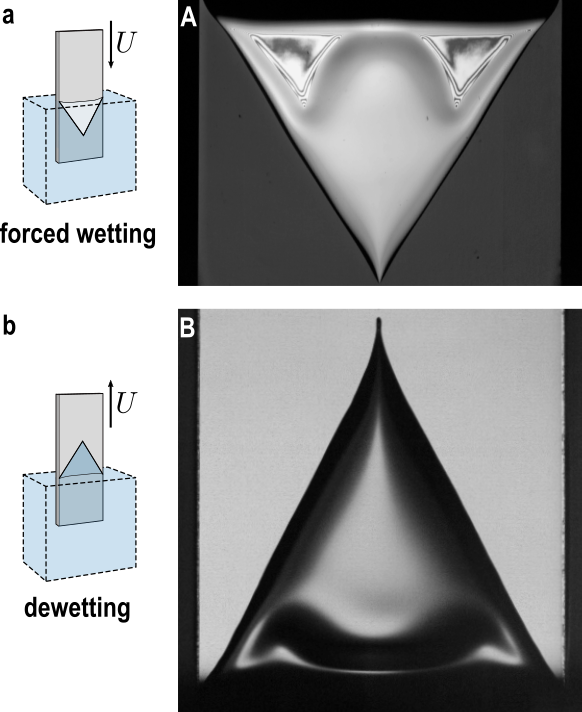}
	\caption{(a) Schematic of forced wetting.  A film of air is entrained into a liquid bath by the surface of a moving substrate.  (A) Front view of steady state of an entrained air film in forced wetting.  A mylar tape of width $w = 12.7$ mm plunges vertically into a water-glycerin mixture of viscosity $\eta\textsubscript{out} = 226$ cP at $U = 130$ mm/s.  The entrained air layer assumes a $V$ shape with two thin, flat sections at the upper corners.  (b) Schematic of dewetting.  A film of liquid is dragged out from a liquid bath by the surface of a moving substrate.  (B) Front view of steady state of a liquid film in dewetting.  An acrylic plate of width $w = 20.3$ mm is pulled vertically out of a water-glycerin bath of viscosity $\eta = 100$ cP at $U =4.4$ mm/s.  An entrained liquid layer forms an upside-down $V$ shape with two thin sections at the lower corners.}
	\label{fig:comparison}
\end{figure}

In ``forced wetting'', a solid substrate rapidly enters a liquid bath with a film of air entrained along its surface (Fig.~\ref{fig:comparison}a). The liquid \emph{wets} the solid.  Conversely, in ``dewetting'', a solid is rapidly withdrawn from a liquid bath, dragging out a film of liquid (Fig.~\ref{fig:comparison}b). The liquid film \emph{dewets} the surface as it is pulled down by gravity.  This paper will show that forced wetting and dewetting share many similarities.  

The phenomenon of forced wetting or dewetting can be observed commonly in daily life.  Yet, it can often be difficult to see the \emph{steady-state} behavior of forced wetting or dewetting, as distinct from capturing only the initial onset.  In order to characterize the long-time limit of forced wetting, we developed a system in our previous study~\cite{he_nagel2019} to maintain the entrained air layer until a steady state is reached.  By steadily  pushing a long ribbon of mylar tape into a liquid, the air layer develops prominent and surprising structures.  Figure~\ref{fig:comparison}A shows that at steady state, the air layer assumes a $V$ shape with two extremely flat and thin sections positioned at the upper corners.

This paper extends this study to the process of dewetting.  Figure~\ref{fig:comparison}B shows that at steady state, the entrained liquid layer forms an upside-down $V$ shape.  Both in the case of dewetting and wetting (as shown in~\cite{he_nagel2019}), two sharply different thicknesses stably coexist inside a triangular-shaped contact line (the solid/liquid/gas interface).  There is a thin-thick alternation of the entrained fluid that appears near the bottom (in dewetting) or near the top (in wetting) across the width of the substrate.  Despite the well-known fundamental difference in advancing and receding contact line motions~\cite{bonn2009}, there is a striking similarity between the structures found in both experiments.

%The phenomenon of forced wetting or dewetting can be observed commonly in daily life.  Yet, in a casual setting, it is difficult to see the \emph{steady-state} behavior of such processes in its intermittent appearances.  The surface properties of the substrate is usually nonuniform, the duration of the event too short and the geometry of the boundaries too irregular.  A controlled experiment can prolong the process to reveal a stationary wetting/dewetting layer that can be quantified with high-speed video techniques.  

The study of wetting/dewetting began long before the age of high-speed imaging.~\cite{ablett1923,wenzel1936}.  Various aspects have been addressed such as  deposited layer thickness~\cite{landau1942,derjaguin1943,white1964,wilson1982},  maximum wetting speed~\cite{Deryaguin1964,wilkinson1975, burley1976, blake1979maximum, burley1984, benkreira2008, benkreira2010}, contact angles~\cite{gutoff1982,sedev1991, petrov1992, marsh1993}, confinement effects~\cite{vandre2012, vandre2014,kim_nam2017}, and the \textit{onset} of the entrainment transition~\cite{eggers2004, asdf2007, asdf2008, snoeijer2006, chan2012, qin2018, kamal2019}.  The main purpose of this paper, on the other hand, is to study the long-time evolution of the contact line motion, and to characterize the prominent structure in the entrained fluid layer at steady state.  By measuring interference fringes as a function of the angle of incidence of the light source, the \textit{absolute} thickness of the wetting layer was determined as a function of different control parameters.

%The measurements reveal simple scaling laws while add complexity to various models of current understanding, suggesting the need for a new perspective for this old problem. 

Forced wetting or dewetting can occur in different geometries~\cite{bretherton1961,taylor1961,deryck1996,zhao2018, gao2019}.  Notably, a series of studies shows that the tail of a sliding droplet is in many aspects an equivalent problem~\cite{podgorski2001,limat2004,legrand_limat2005,rio2005,snoeijer_limat2005,snoeijer_eggers2007,peters2009,winkels2011,limat2014}.  The observations and conclusions presented in the present work may suggest similar behavior in those situations and contribute a new perspective for wetting/dewetting in its various other forms.

\section{Experiments}
\subsection{Mechanical apparatus, fluids and substrate preparation}
\begin{figure}[htp]
	\includegraphics[width=1\textwidth]{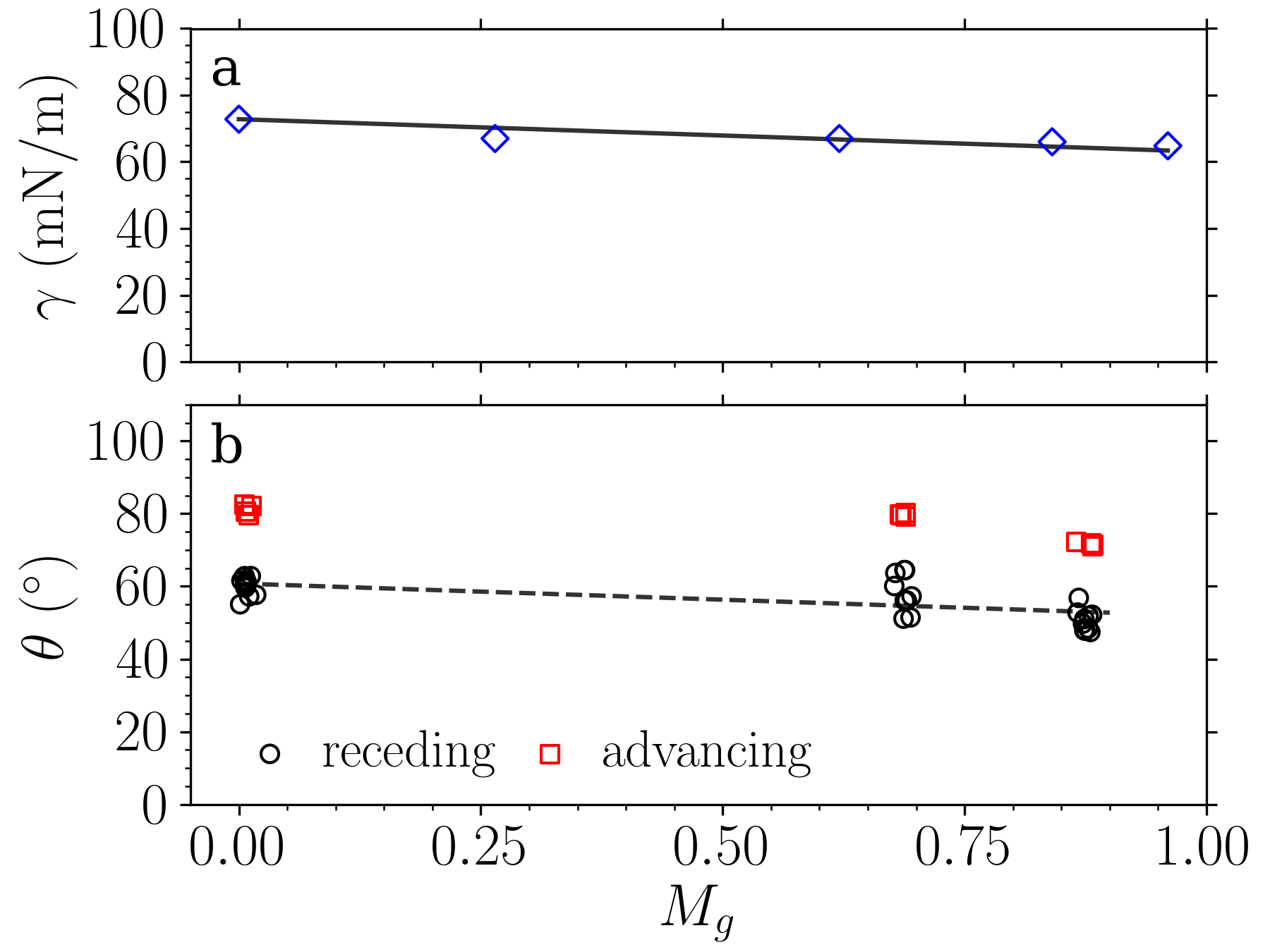}
	\caption{(a) Measured liquid-air interfacial tension $\gamma$ versus glycerin mass fraction $M_g$ of water-glycerin mixtures.  Solid line: prediction of Eq.~\ref{eq:fraction}.  (b) Receding and advancing contact angles on an acrylic substrate (Rain-X$\textsuperscript{\textregistered}$ coated) versus $M_g$.  At each value of $M_g$, the data is displaced slightly in the x-direction to make overlapping symbols visible.  Dashed line: linear fit for receding contact angles.}
	\label{fig:gamma_CA}
\end{figure}

For the dewetting experiments, the solid substrate was installed on a stage on a vertical linear guide (PBC Linear $\text{\texttrademark}$, IVTAAW).  The system was driven by a step motor (Silverpak 17) through a roller chain.  The substrate velocity ranged from 1 mm/s to 400 mm/s with $<$5$\%$ fluctuation.  Water-glycerin mixtures were used as the viscous liquid with a viscosity range 0.9 cP $\leq\eta\leq$ 1264 cP.  The viscosity was measured by an Anton Paar MCR301 rheometer or by manual glass viscometers (CANNON-Ubbelohde).  The density $\rho$ and the liquid-air interfacial tension $\gamma$ of the mixture were measured by a KR\"USS tensiometer, which were consistent with literature values~\cite{glycerine_overview, glycerine_tension}.  The measured $\gamma$'s are plotted in Fig.~\ref{fig:gamma_CA}a, where the solid line shows that they can be well described by the fractional contribution model~\cite{glycerine_tension}:   
\begin{align}
	\gamma = M_g\gamma_g + (1-M_g)\gamma_w,
	\label{eq:fraction}
\end{align}
where $M_g$ is the mass fraction of the glycerin in the mixture, $\gamma_g$ the surface tension of pure glycerin and $\gamma_w$ the surface tension of pure water.

In the dewetting experiments, the solid substrate consists of slender rectangular sections (560 mm$\times$3.2 mm) of black cast acrylic, cut to various widths, 12.3 mm $\leq w \leq$ 50.5 mm.  The edges were further milled and polished to prevent contact-line pinning during the experiments.  The substrate surface was first wiped with isopropyl alcohol to remove chemical residues from the manufacturing process.  Rain-X$\textsuperscript{\textregistered}$ Original Glass Water Repellent (PDMS) was then applied to the surface and then wiped off.  The static contact angles were measured from side-view images of the drops of the mixtures against the prepared acrylic substrate, shown in Fig.~\ref{fig:gamma_CA}b~\footnote{Side-view images provided by Chloe W. Lindeman}.  The advancing values (red squares) were measured from the drops soon after they were deposited onto the substrate.  To obtain the receding values (black circles), the contact line of a drop was made to move at a negligible velocity ($<50$ $\mu$m/s) by continuously adding or removing liquid through a needle, and the data scatter in Fig.~\ref{fig:gamma_CA}b reflects the fluctuations of the stick-slip motion.  Following Le Grand \textit{et al}~\cite{legrand_limat2005}, circles were fitted locally whose tangent lines were found $\sim30$ $\mu$m near the contact lines to ensure reproducibility of the angle measurements.  As can be seen in Fig.~\ref{fig:gamma_CA}b, the gaps between the advancing and receding contact angles show a significant contact angle hysteresis $\sim20^{\circ}$ of water-glycerin mixtures on the Rain-X$\textsuperscript{\textregistered}$ coated acrylic.  A linear fit (dashed line) gives an empirical formula:
\begin{align}
	\theta_r = 60.78^{\circ}-8.9^{\circ}M_g.
	\label{eq:theta_r}
\end{align}

Both $\gamma$ and $\theta_r$ are nearly constants in our experimental range, the weak decreasing trend of which can be sufficiently characterized by the linear approximations Eqs.~\ref{eq:fraction} and~\ref{eq:theta_r}.  As will be shown in Section~\ref{sec:umax}, $\gamma$ and $\theta_r$ have limited impact on the data analysis. 

For the forced-wetting experiments, commercial magnetic mylar tape (cassette tape with $w=$ 6.4 mm, VHS tape with $w=$ 12.7 mm and recording tape with $w=$ 25.4 mm) was used as 
the substrate.  Water-glycerin mixture were used as the outer fluid with viscosity 26 cP $\leq \eta\textsubscript{out}\leq$ 572 cP.

Measurement of refractive index (consistent with literature~\cite{hoyt1934}) and the thickness of the entrained liquid layer are described below and in Ref.~\cite{he_nagel2020optics}.

\subsection{Measurement of absolute film thickness}
Haidinger's fringes, to be distinguished from Newton's fringes, refer to the interference pattern produced by a varying angle of incidence, $\theta$, of the light source~\cite{rayleigh1906,raman1940}.  To obtain the \textit{absolute} thickness, $h$, of an entrained fluid layer, I have developed an interferometric method based on measuring Haidinger's fringes given by the sample.  $h$ affects how the optical path difference, $\Delta L$, changes with $\theta$.  By measuring $\Delta L$ (or equivalently, the interference intensity) as a function of $\theta$ at at given point, $h$ at that point can be deduced.  The measuring device is similar to one used to measure thickness of oxide layers and silicon semiconductor samples~\cite{gold1991,kim2018}.  The derivation of the principle is shown in Appendix A and new features and capabilities of this method are described in Ref.~\cite{he_nagel2020optics}.

To map the detected interference pattern to sample thickness $h$, I have developed a new algorithm using likelihood maximization (Appendix B).  This algorithm facilitates a reliable pattern detection and a precise topography reconstruction.  

Combining the above techniques, I am able to identify precisely the absolute thickness of the entrained fluid layer systematically.

\section{Results}
\subsection{Formation of a $V$-shaped structure}
\begin{figure}[htp]
	\includegraphics[width=1\textwidth]{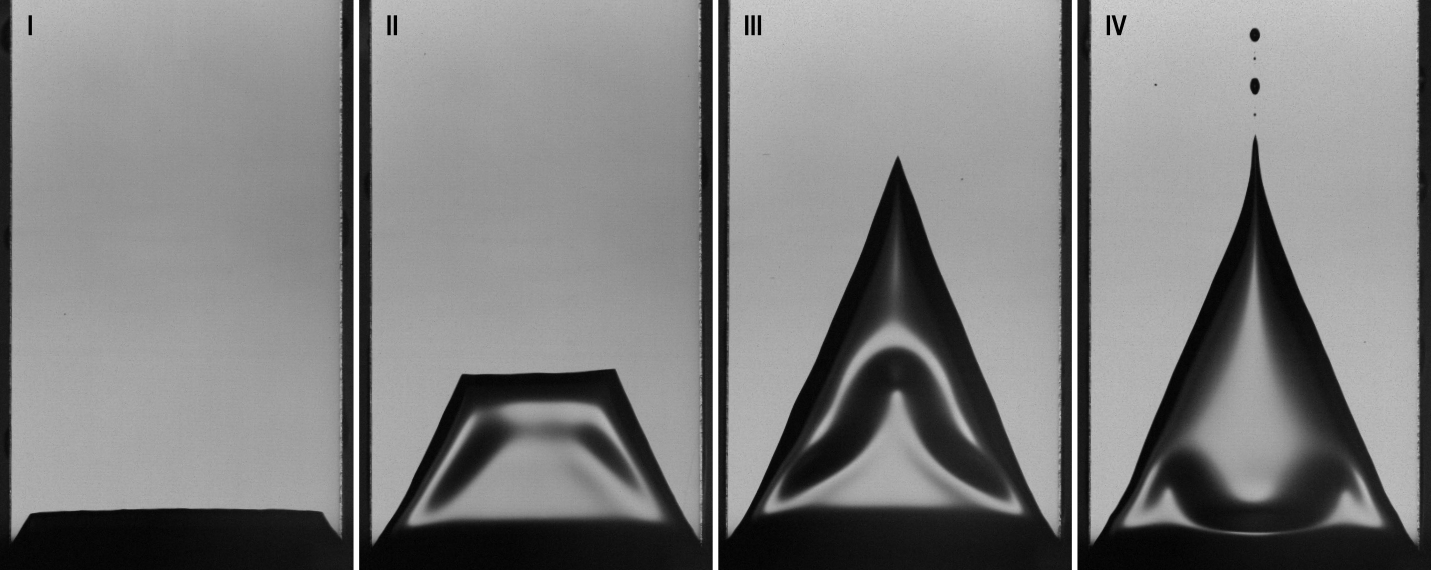}
	\caption{Images of a typical evolution of a dewetting liquid layer, for a duration $\approx$ 40 s.  An acrylic plate of width,  $w=20.3$ mm, is pulled vertically out of a water-glycerin bath of viscosity $\eta=100$ cP at $U=$ 4.4 mm/s.  The fluid layer reaches its steady state in frame IV.  (The images I-IV are not evenly spaced in time in order to show the different stages clearly.)}
	\label{fig:formation}
\end{figure}

Figure~\ref{fig:formation} shows a series of images of a fluid being pulled out of a bath by a flat substrate with straight edges.  The substrate travels vertically upward at a constant velocity while the liquid forms a thin layer entrained to its surface.  Initially, as shown in the first frame (I), the contact line rises from the bath in the form of a trapezoid composed of a central nearly horizontal section with two short sloping sides.  In the second frame (II), the trapezoid grows in height.  The sides remain at the same angle and the central horizontal section advances upward and becomes narrower.  Just behind the contact line there is a thick ridge; further back there is an extended thin, flat region.  In the third frame (III) the horizontal contact line moves upward until the trapezoid closes into a triangle.  At this point, the central part of the thick ridge starts to widen and spread downward.  In the final frame (IV), as the system reaches its steady-state shape, the thin part is split into two smaller sections at the lower corners of the entrained fluid.  

There appears a thin-thick alternation in the spanwise direction near the bottom of the entrained layer.  In steady state, as shown in the fourth image (IV), the triangular contact line retracts slightly so that the tilted sides are less steep compared to its shape during formation shown in the third frame.  When the velocity of the substrate is large enough, small liquid drops, attached to the moving substrate, emerge at the tip of the triangular pocket, also shown in the last frame (IV).  

\onecolumngrid

\twocolumngrid

\begin{figure}[htp]
	\includegraphics[width=0.7\textwidth]{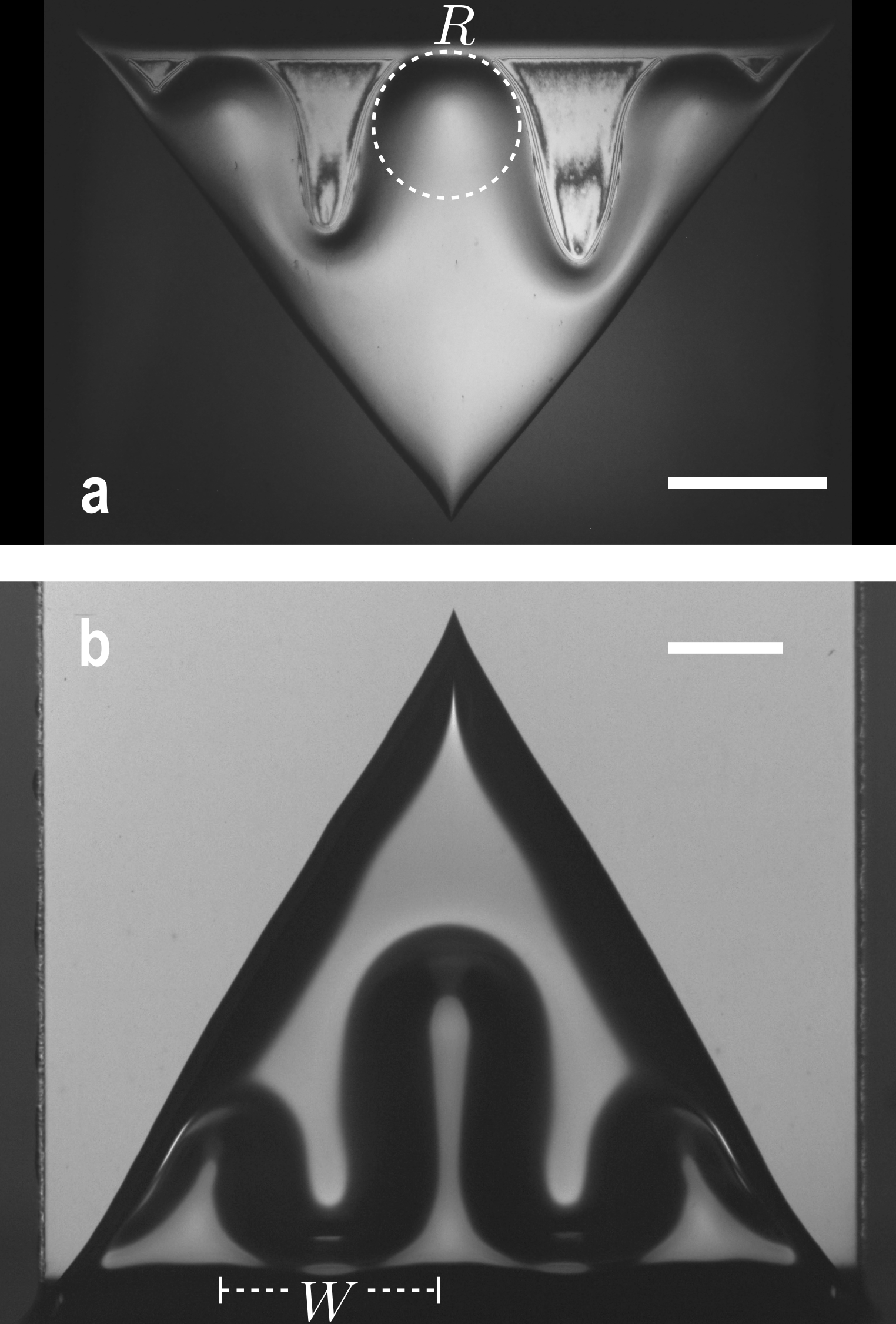}
	\caption{Increasing substrate width, $w$, increases the number of thin-thick alternations in film thickness.  (a) A 25.4 mm wide recording tape entrains an air film with 3 thick sections.  The upper portion of the thin-thick boundary can be approximated by a circle of radius $R$.  (b) A 35.3 mm wide acrylic plate entrains a water-glycerin film with 2 thick sections with width $W$.  In a and b the substrate widths, $w$, are twice what they were in  the corresponding images shown in Fig.~\ref{fig:comparison}a and b.  Scale bar: 5mm.}
	\label{fig:thickwidth}
\end{figure}

The number of alternating undulations of the layer thickness in the transverse direction depends on the width of the substrate.  Figure~\ref{fig:thickwidth}a shows steady state of a forced wetting film on a mylar surface of width $w=$ 25.4 mm.  Compared to Fig.~\ref{fig:comparison}a (substrate width $w=$ 12.7 mm), Fig.~\ref{fig:thickwidth}a shows that near the top of the $V$ there are 4 thin sections.  The thin-thick features are more extended for wider substrates than they are for a narrower ones. 

The same trend applies for the case of dewetting.  Fig.~\ref{fig:thickwidth}b shows the dewetting film contains multiple marked thin-thick alternations for a $w=$ 35.3 mm wide acrylic plate.

\begin{figure}[htp]
	\includegraphics[width=1\textwidth]{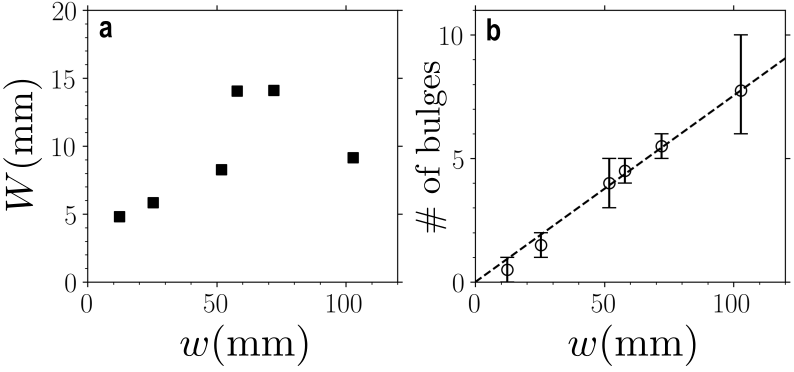}
	\caption{Geometry of dewetting films.  (a) Width of a thick section, $W$, versus substrate width, $w$, for water dewetting a mylar surface.  (b) Number of thick parts, $N$, versus $w$, for water dewetting a mylar surface.  Error bars indicate different thin-thick configurations observed, for a given $w$.  Dashed line: linear fit to the mean observation.}
	\label{fig:thickfingers}
\end{figure}

To quantify this trend, I measured in dewetting the width of the thick bulge $W$ as indicated by the dashed line in Fig.~\ref{fig:thickwidth}b.  Fig.~\ref{fig:thickfingers}a shows that this width do not remain a constant with increasing substrate width.  For wide substrates, $w>$ 70 mm in the case of water dewetting on mylar, the adjacent thick parts constantly merge and re-split, so the width of a single bulge fluctuates considerably.  Despite these fluctuations, Fig.~\ref{fig:thickfingers}b shows that there is linear relationship between the number of thick bulges versus the substrate width, $w$.  

\begin{figure}[htp]
	\includegraphics[width=0.8\textwidth]{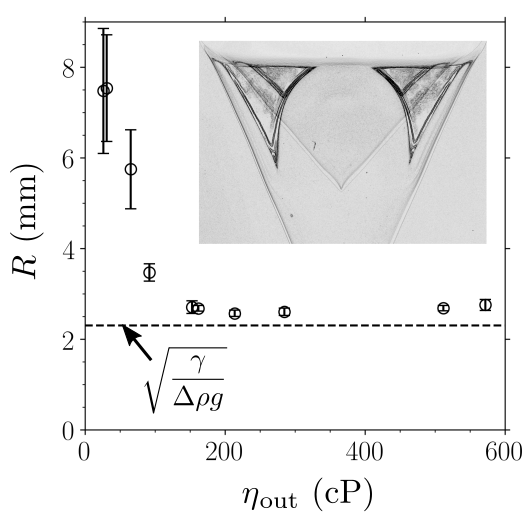}
	\caption{Geometry of forced-wetting films.  Radius of curvature $R$ of the upper portion of the thick part is plotted against liquid viscosity $\eta\textsubscript{out}$.  Dashed line: capillary length $l_c=\sqrt{\gamma/\Delta \rho g}$.  Inset: superimposed edge-detected images of two different air films, entrained into different liquids of viscosities $\eta\textsubscript{out}=$ 160 cP and 290 cP respectively at $U=$ 150 mm/s.}
	\label{fig:RvsMu}
\end{figure}

The front-view shape of the thick bulges in forced wetting, on the other hand, is very well defined.  A large portion near the tip of the thick section can be approximated by a circle, as indicated in Fig.~\ref{fig:thickwidth}a.  In Fig.~\ref{fig:RvsMu} I have plotted the measured radius, $R$, of the fitted circle, versus liquid viscosity $\eta\textsubscript{out}$, for a fixed substrate width $w=$ 12.7 mm.  The error bars indicate the influence on $R$ of different substrate velocities.  The radius decreases with an increase of the liquid viscosity, and saturates at a value close to the capillary length $l_c = \sqrt{\gamma/\Delta\rho g}$, independent of the substrate velocity.  This suggests that at a large viscosity ($\eta\textsubscript{out}>$ 90 cP) the shape of the thin-thick boundary of the air layer is selected by a balance between buoyancy and interfacial tension.  The inset of Fig.~\ref{fig:RvsMu} shows the superimposed images of two air layers entrained in two different viscous liquids (both with $\eta\textsubscript{out} >$ 90 cP).  The two shapes have very different $V$-shaped outlines, but the curves of the thin-thick boundary overlap very well.

\subsection{Wetting velocity of the contact line}\label{sec:umax}
The upside-down $V$ shape of the contact line was first noted by Deryaguin and Levi~\cite{Deryaguin1964}, and was quantitatively interpreted by Blake and Ruschak in terms of the maximum velocity that a contact line could move across a substrate~\cite{blake1979maximum}.  When the substrate velocity $U$ exceeds the maximum value, $U\textsubscript{max}$, the stationary contact must become inclined by an angle $\phi$ (with respect to the horizontal direction) so that the velocity of the contact line normal to its surface, $U_n$, remains at a constant value $U\SB{max}$:

\begin{align}
	\label{eq:umax}
	U_n = U \cos\phi \le U\textsubscript{max} .
\end{align}

In this section I will show a few experimental observations on contact line movement that imply more complexity than a simple model of a constant $U_n$.

\begin{figure}[H]
	\includegraphics[width=0.92\textwidth]{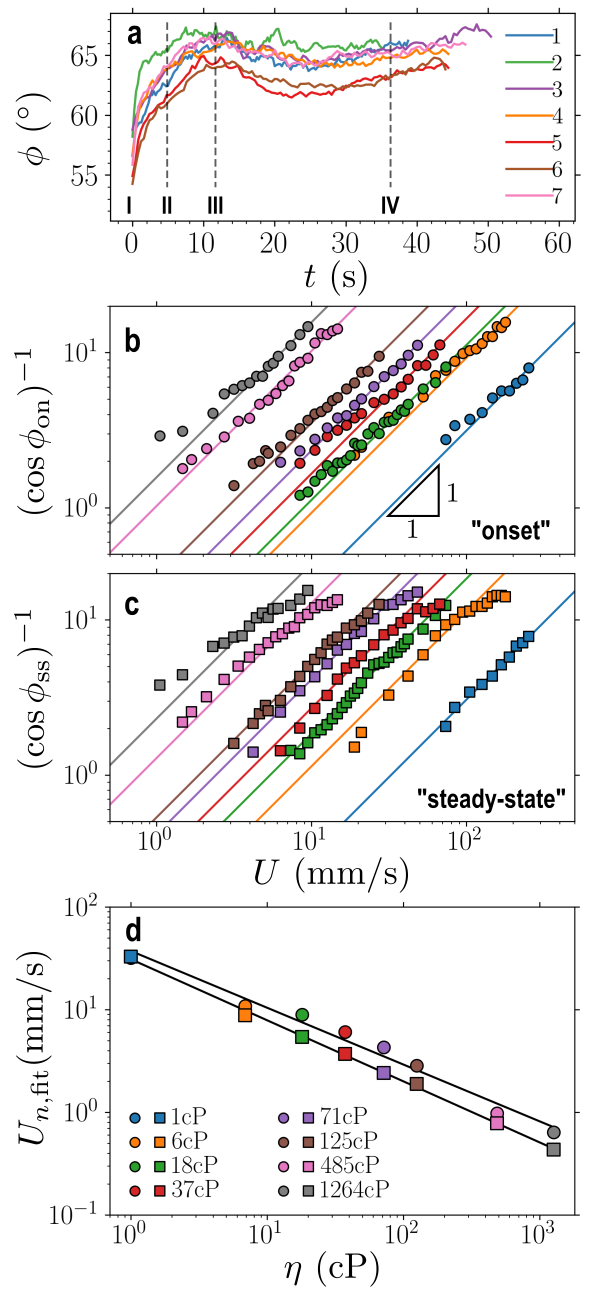}
	\caption{Contact-line inclination and contact-line velocity in dewetting.  (a) Evolution of the inclination angle $\phi$ (relative to the horizontal direction) of the lateral contact lines versus time $t$ for 7 different runs under the same condition.  Plate velocity $U=4.4$ mm/s.  Liquid viscosity $\eta=100$ cP.  (b) and (c) $(\cos\phi)^{-1}$ versus $U$ for onset ($\phi\textsubscript{on}$, circles) and steady state ($\phi\textsubscript{ss}$, squares) respectively. Solid lines:  linear fits to each data set (slope=1).  (d) Fitted normal relative velocity, $U_n\textsubscript{,fit}$, versus liquid viscosity $\eta$ for both onset (circles) and steady state (squares).  Top solid line: $U_n\textsubscript{,fit}\propto\eta^{-0.55}$.  Bottom solid line: $U_n\textsubscript{,fit}\propto\eta^{-0.59}$.  Legend is consistent for (b), (c) and (d).}
	\label{fig:angles}
\end{figure}

For a typical experimental condition ($\eta=100$ cP, $U=4.4$ mm/s, $w=20.3$ mm),  Fig.~\ref{fig:angles}a shows the inclination angle $\phi$ (relative to the horizontal direction) of the lateral contact line, tracked continuously over time $t$ for 7 repeated runs.  Clearly $\phi$, and therefore $U_n$ from Eq.~\ref{eq:umax}, do not remain constant during the formation of the $V$ shape;  An obvious trend can be observed: the inclination first assumes an onset value $\phi\textsubscript{on}$, reaches a maximum, and plateaus at a steady state value $\phi\textsubscript{ss}$.  There are considerable fluctuations from run to run, but the variations do not show a time dependence (the order of runs is shown in the legend), ensuring no apparent ``aging'' effect of the Rain-X$\textsuperscript{\textregistered}$ coated substrate.  A rough correspondence to the fluid shape imaged in Fig.~\ref{fig:formation}I-IV is also noted in Fig.~\ref{fig:angles}a.  The maximum values of $\phi$ consistently occur at the completion of the triangular pocket shape (Fig.~\ref{fig:formation}III).  Notice the extended duration of the experiments ($\sim40$s) to reach steady state, which is much longer than that of forced wetting ($\sim300$ms, see Fig.~1 of~\cite{he_nagel2019}), and often makes observation of steady state difficult in dewetting experiments.

Since $\phi\textsubscript{on}$ is consistently smaller than $\phi\textsubscript{ss}$, it is necessary to conduct separate analyses on both of these quantities.  Motivated by Eq.~\ref{eq:umax}, I plot in Fig.~\ref{fig:angles}b, c $(\cos\phi)^{-1}$ versus $U$ at onset and in steady state respectively, for various liquid mixtures.  The linear trends indicate that there exists a constant $U_n$ independent of $U$ for each viscosity, but the value is different at onset from what it is in steady state.  The data for steady state in Fig.~\ref{fig:angles}c depart significantly from linearity at larger velocities, possibly suggesting a different regime that I will not focus on in the present work.  

I fit the data to the relation $(\cos\phi)^{-1} = U_{n,\text{fit}}^{-1}U$ (Eq.~\ref{eq:umax}) for each liquid mixture.  The resulting $U_{n,\text{fit}}$ is shown in Fig.~\ref{fig:angles}d, plotted versus $\eta$.  The average normal velocity $U_{n,\text{fit}}$ is consistently larger at onset than it is in steady state as was noted above for the special case in Fig.~\ref{fig:angles}a.  

Observe that there is an approximate linear relationship in the log-log plot of Fig.~\ref{fig:angles}d, which motivates a power-law fit $U_{n,\text{fit}}\propto\eta^{\alpha}$: 
\begin{subequations}
\begin{align}
	\label{eq:power_on}
	U_{n,\text{fit}; \text{on}}\propto \eta^{ -0.55 \pm 0.03},\\ 
	\label{eq:power_ss}
	U_{n,\text{fit}; \text{ss}}\propto \eta^{ -0.59 \pm 0.01}.
\end{align}
\end{subequations}
Incorporating the surface tension $\gamma$ or the static contact angle $\theta_r$ in the regression as $U_{n,\text{fit}}\propto\eta^{\alpha}\gamma^{\beta}$ or $U_{n,\text{fit}}\propto\eta^{\alpha}\theta_r^{\beta}$ shows that $\beta$ is insignificantly different from 0~\footnote{$p$-value $>0.05$; also notice $\gamma$ and $\theta_r$ are linearly related through Eq.~\ref{eq:fraction} and~\ref{eq:theta_r} so there is no need to further test $U_{n,\text{fit}}\propto\eta^{\alpha}\gamma^{\beta}\theta_r^{\gamma}$.}.  The simple empirical laws Eqs.~\ref{eq:power_on} and~\ref{eq:power_ss} well describe the data for over 3 decades of viscosity $\eta$, as shown by the solid lines in Fig.~\ref{fig:angles}d.

\begin{figure}[htp]
	\includegraphics[width=1\textwidth]{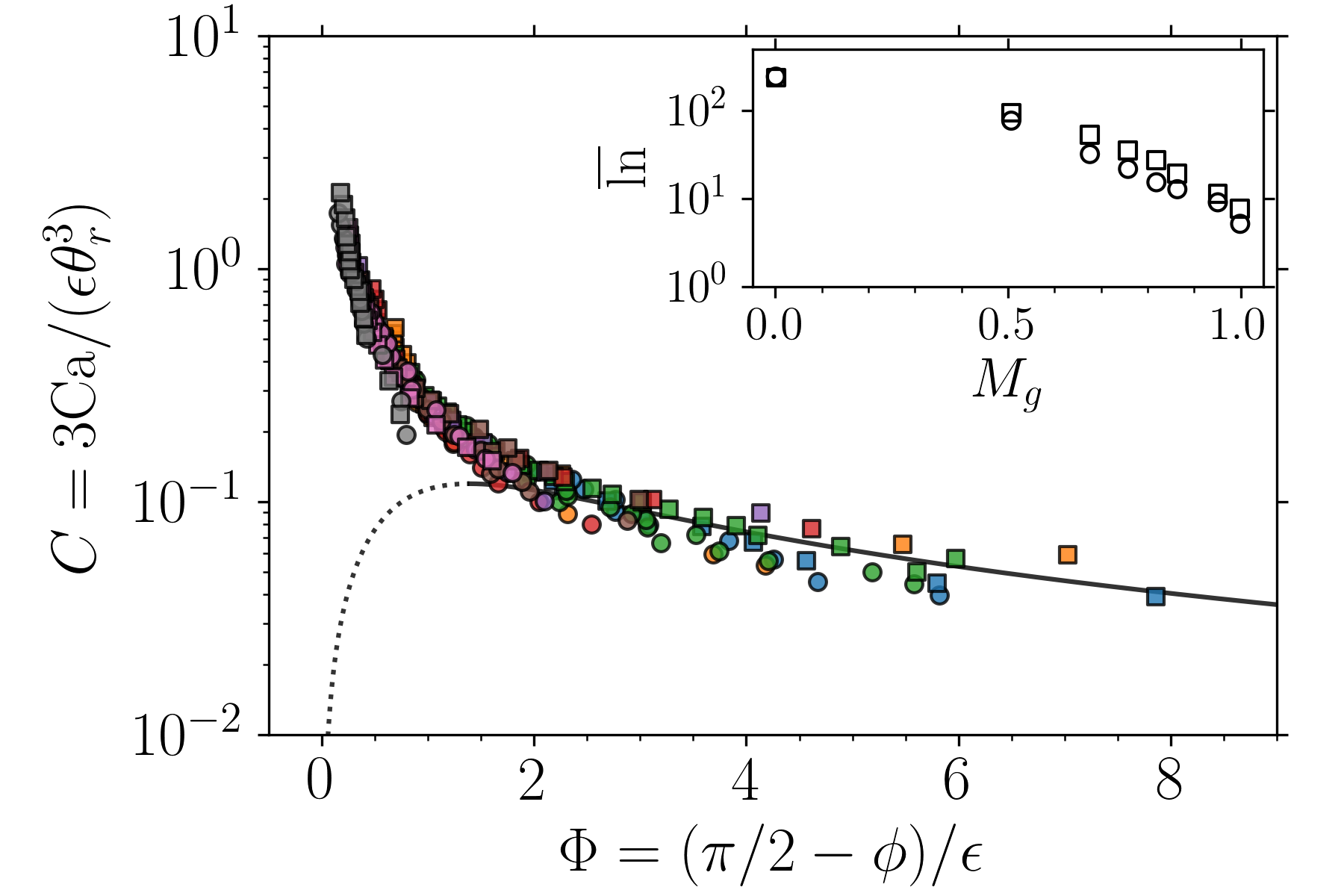}
	\caption{Data collapsing for dewetting using Eq.~\ref{eq:snoeijer_corner}.  Legend consistent with Fig.~\ref{fig:angles}b,c and d.  Dotted line: unstable brance of Eq.~\ref{eq:snoeijer_corner}.  Solid line: stable branch.  Inset: fitted $\overline{\mathrm{ln}}$ using Eq.~\ref{eq:voinov_cox_degennes} versus the glycerin mass fraction $M_g$.}
	\label{fig:C_Phi}
\end{figure}

Yet it should be emphasized that Eqs.~\ref{eq:power_on} and~\ref{eq:power_ss} are only phenomenological.  The maximum wetting speed $U_{n}$ may depend not only on $\eta$, $\gamma$ and $\theta_r$ but also the microscopic quantities such as the slip length $l$.  Here I shall examine the simple model proposed by Le Grand \textit{et al}~\cite{legrand_limat2005}:
\begin{align}
	\label{eq:voinov_cox_degennes}
	U_{n} = \frac{\gamma}{\eta}\frac{\theta_r^3-(\theta_r/\sqrt{3})^3}{9\overline{\mathrm{ln}}}, 
\end{align}
where $\overline{\mathrm{ln}}\equiv \mathrm{ln}(x/l)$ is the logarithmic ratio of the macroscopic/microscopic scales.  It is a combination of the classical Cox-Voinov relation~\cite{voinov1976, cox1986} $U_n \propto \theta_r^3-\theta_m^3$ where $\theta_m$ is the receding contact angle at $U_n$, and De Gennes' model~\cite{degennes1986} $\theta_m = \theta_r/\sqrt{3}$.  Note Eq.~\ref{eq:voinov_cox_degennes} is not in immediate contradiction to Eq.~\ref{eq:power_on} and ~\ref{eq:power_ss}, because $\eta$, $\gamma$, $\theta_r$ and $\overline{\mathrm{ln}}$ may all be related through the mixture fraction $M_g$.  Since the slip length $l$ cannot be directly measured from the current setup, $\overline{\mathrm{ln}}$ was used as a fitting parameter for Eq.~\ref{eq:voinov_cox_degennes}, and the fitted result is shown in the inset of Fig.~\ref{fig:C_Phi}.  It turns out that for the pure glycerol $M_g=1$, the slip length $l=xe^{-\overline{\mathrm{ln}}}\sim10^{-8}$ m is indeed in the nanometer scale corresponding to the molecular size.  However, for all other water-glycerin mixtures, including pure water, $\overline{\mathrm{ln}}$ is unusually large and as a result $l \ll 10^{-9}$ m is unphysically small.  Similar discrepancies have been found in experiments of sliding droplets of water~\cite{podgorski2001, winkels2011}.

Snoeijer \textit{et al} matched the contact line dyanmics with the similarity solution of a corner flow (Limat-Stone model~\cite{limat2004, snoeijer_limat2005}) for a sliding droplet, and found the relation~\cite{snoeijer_eggers2007}:
\begin{align}
	C = \frac{6\Phi}{35+18\Phi^2},
	\label{eq:snoeijer_corner}
\end{align}
where $C=3\mathrm{Ca}/(\epsilon\theta_r^3)$, $\Phi=(\pi/2-\phi)/\epsilon$, and $\epsilon=1/\sqrt{\overline{\mathrm{ln}}}$.  Using the above fitting result of $\overline{\mathrm{ln}}$, despite the lack of its physical interpretation, the data in Fig.~\ref{fig:angles} can also be collapsed by Eq.~\ref{eq:snoeijer_corner}, shown in Fig.~\ref{fig:C_Phi}.  In the theory of Eq.~\ref{eq:snoeijer_corner}, the stable branch only starts at $\Phi=\sqrt{35/18}\approx1.4$ (solid line). In the unstable branch (dotted line) a rivulet solution takes place where the liquid is continuously deposited through a thin stream (which would further break into sessile droplets) at the tip of the $V$, as observed in Fig.~\ref{fig:formation}IV.  It is difficult to quantitatively verify the onset of the rivulet solution at $\Phi\approx1.4$ here using the current imaging resolution and frame size for the $V$ shape.  Interestingly, the data collapse works for the whole range of $\Phi$, for both the supposed corner regime (solid line) and the rivulet regime (dotted line).  I find that the data collapse as well as the size of $\overline{\mathrm{ln}}$ do not depend too critically on the model of Eq.\ref{eq:voinov_cox_degennes}, although it indeed gave a somewhat better collapse compared with several other hydrodynamic variations (Voinov and Cox~\cite{voinov1976, cox1986}, Dussan~\cite{dussan1979}, De Gennes and Brochard-Wyart~\cite{degennes1986, brochardwyart1991} and Eggers~\cite{eggers2004}).  In the meantime,  I shall point out that Eq.~\ref{eq:snoeijer_corner} is not trivially equivalent to Eq.~\ref{eq:voinov_cox_degennes} (only if $18\Phi^2\gg 35$~\cite{snoeijer_limat2005}) for my experimental regime ($18\Phi^2$ changes from 2 to 200) so that the collapse is not automatic.  Curiously, Eq.~\ref{eq:snoeijer_corner} was based on a self-similar flow without gravity effect for a sliding droplet, yet works for plate dewetting here of a more complex geometry (thin-thick alternation) and a larger length scale, suggesting that the underlining dynamics may be closely related.

For the case of steady-state forced wetting, our previous work~\cite{he_nagel2019} concluded a power law for the relation between the normal wetting velocity,  $U_n$,  versus the viscosity of the \textit{outer} liquid, $\eta\textsubscript{out}$, where an air pocket is entrained into:
\begin{align}
	\label{eq:umax_wetting}
	U_n \propto \eta\textsubscript{out}^{-0.75\pm 0.03}.
\end{align}
Therefore, the normal relative velocity, $U_n$, decreases with increasing viscosity both in the case of the inner fluid (as in the case of dewetting) and for the outer fluid (as in wetting).  
%Since the fitted exponent is larger (absolute value) in Eq.~\ref{eq:umax_wetting}, changing the outer fluid %has a larger impact on the normal wetting velocity of the contact line.    

\begin{figure}[htp]
	\includegraphics[width=1\textwidth]{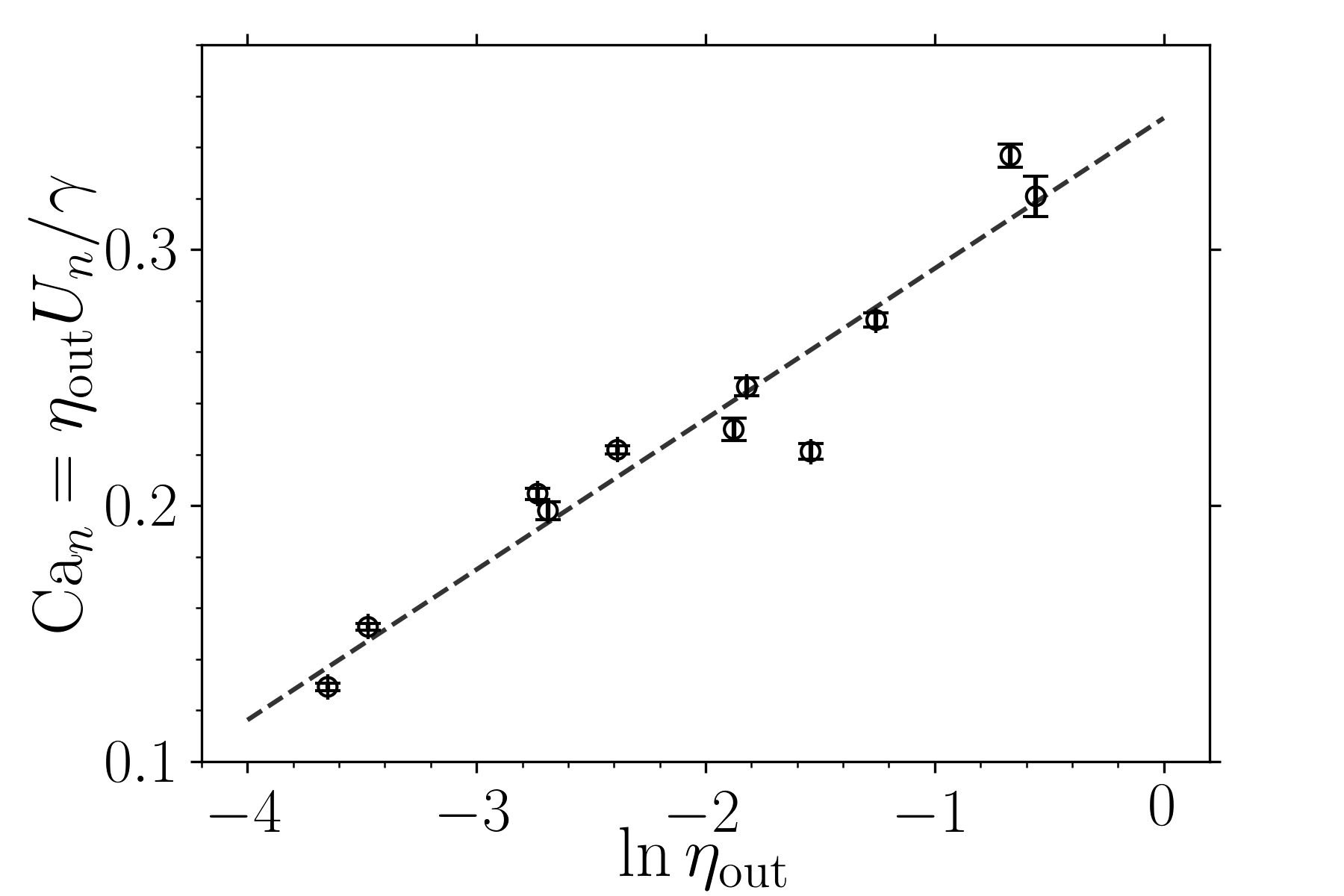}
	\caption{Forced wetting data plotted as $\mathrm{Ca}_n$ versus $\ln\eta\textsubscript{out}$.  Solid line: a least-square fit shows a good linear relation.}
	\label{fig:compare_powerlaw}
\end{figure}

Kamal \textit{et al.}~\cite{kamal2019} studied the contact-line motion where the inner and outer fluids have comparable contributions to the dynamics.  This is the case for forced wetting near the contact line.  Although air has a much smaller viscosity than the viscous liquid, its dissipation cannot be neglected because of the sharp wedge near the contact line(see also~\cite{huh_scriven1971, marchand2012,qin2018}).  In their theoretical work they concluded a logarithmic behavior:
\begin{align}
	\label{eq:umax_wetting_theory}
	\mathrm{Ca}_n\equiv\frac{\eta\textsubscript{out}U_n}{\gamma} = C_1\ln \eta\textsubscript{out} + C_2,
\end{align}
where $C_1$, $C_2$ are constants depending on the model details.  To compare with Eq.~\ref{eq:umax_wetting_theory}, I plotted $\mathrm{Ca}_n$ versus $\ln \eta\textsubscript{out}$ in Fig.~\ref{fig:compare_powerlaw} using the same data leading to the empirical relation Eq.~\ref{eq:umax_wetting}.  The linear relationship in Fig.~\ref{fig:compare_powerlaw} indicates Eq.~\ref{eq:umax_wetting} and Eq.~\ref{eq:umax_wetting_theory} are compatible, and our data in forced wetting verifies the logarithmic trend~\footnote{The coefficient $C_1$ given by the linear fit in Fig.~\ref{fig:compare_powerlaw} seems to give an overly small $q\in[-0.5,0.1]$, where $q$ is the exponent of the modelled interface profile $h(x)$ at the contact line $h\sim x^{1/2}+x^{q}$.}. 

\begin{figure}[htp]
	\includegraphics[width=0.85\textwidth]{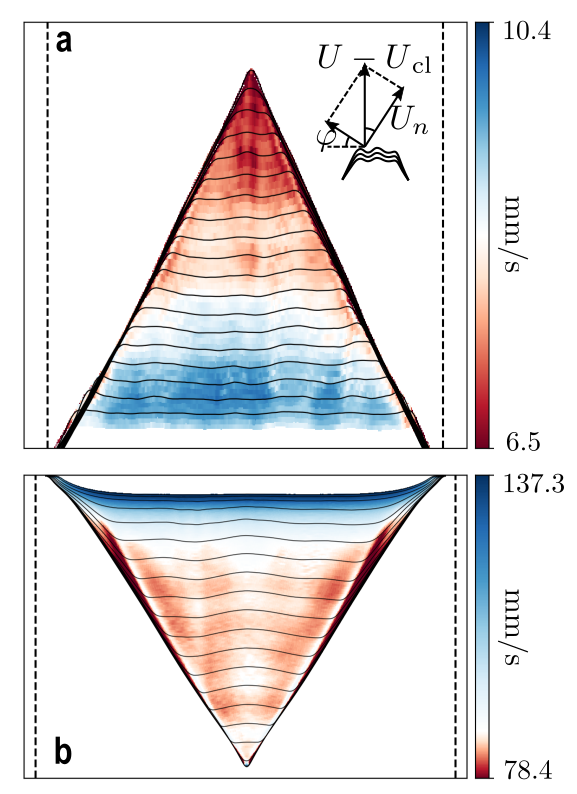}
	\caption{Normal relative velocity, $U_n$, of contact line during formation of a $V$.  Color scale: $U_n$ deduced from Eq.~\ref{eq:un}.  (a) Contact-line positions at intervals of 100 ms, for an acrylic moving out of a water-glycerin mixture of viscosity $\eta=18$ cP at $U=$ 15.8 mm/s.  (b) Contact-line positions at intervals of 50 ms, for a mylar moving into a water-glycerin mixture of $\eta\textsubscript{out}=$ 226 cP at $U=$ 130 mm/s.}
	\label{fig:Ucl}
\end{figure}

Finally I show a visualization of the transition of the velocity of the nearly horizontal contact line, by continuously tracking the contact-line position through a high-speed camera.  For a typical dewetting process, Fig.~\ref{fig:Ucl}a shows the superimposed positions of the contact line at equal time intervals until the trapezoid has reached the triangular shape (as seen in Fig.~\ref{fig:formation}III).  The normal relative velocity $U_n(x,y)$ can then be calculated at each point $(x,y)$ at the contact line (illustrated in the inset of Fig.~\ref{fig:Ucl}a):
\begin{align}
	\label{eq:un}
	U_n(x,y) = [U-U\textsubscript{cl}(x,y)]\cos\varphi(x,y),
\end{align}
where $U$ is the plate velocity, $U\textsubscript{cl}$ is the vertical velocity of a contact-line element at $(x,y)$, and $\varphi(x,y)$ is the \textit{local} inclination of the contact line.  In Fig.~\ref{fig:Ucl}a, the calculated magnitude field $U_n(x,y)$ swept out by the nearly horizontal contact line during this period is mapped in the same figure to a color scale.  The color map shows a significant decrease in normal relative velocity, $U_n$, throughout the process, which is consistent with the above fitted result $U_{n,\text{fit};\text{on}}>U_{n,\text{fit};\text{ss}}$ for the lateral contact lines.  The same conclusion applies for the case of forced wetting, as shown in Fig.~\ref{fig:Ucl}b.  Note that a velocity change of the horizontal contact line soon after entrainment has been observed and modelled with a quasi-steady lubrication theory~\cite{asdf2007, asdf2008}.  The current work further extends the observation till steady state,  which shows more complexity than earlier models/observations of a fixed contact-line velocity for various geometries of dewetting (see e.g.~\cite{redon1991, brochardwyart1991, maleki2007}).

\subsection{Thickness structure of wetting layer}

\begin{figure}[htp]
	\includegraphics[width=0.9\textwidth]{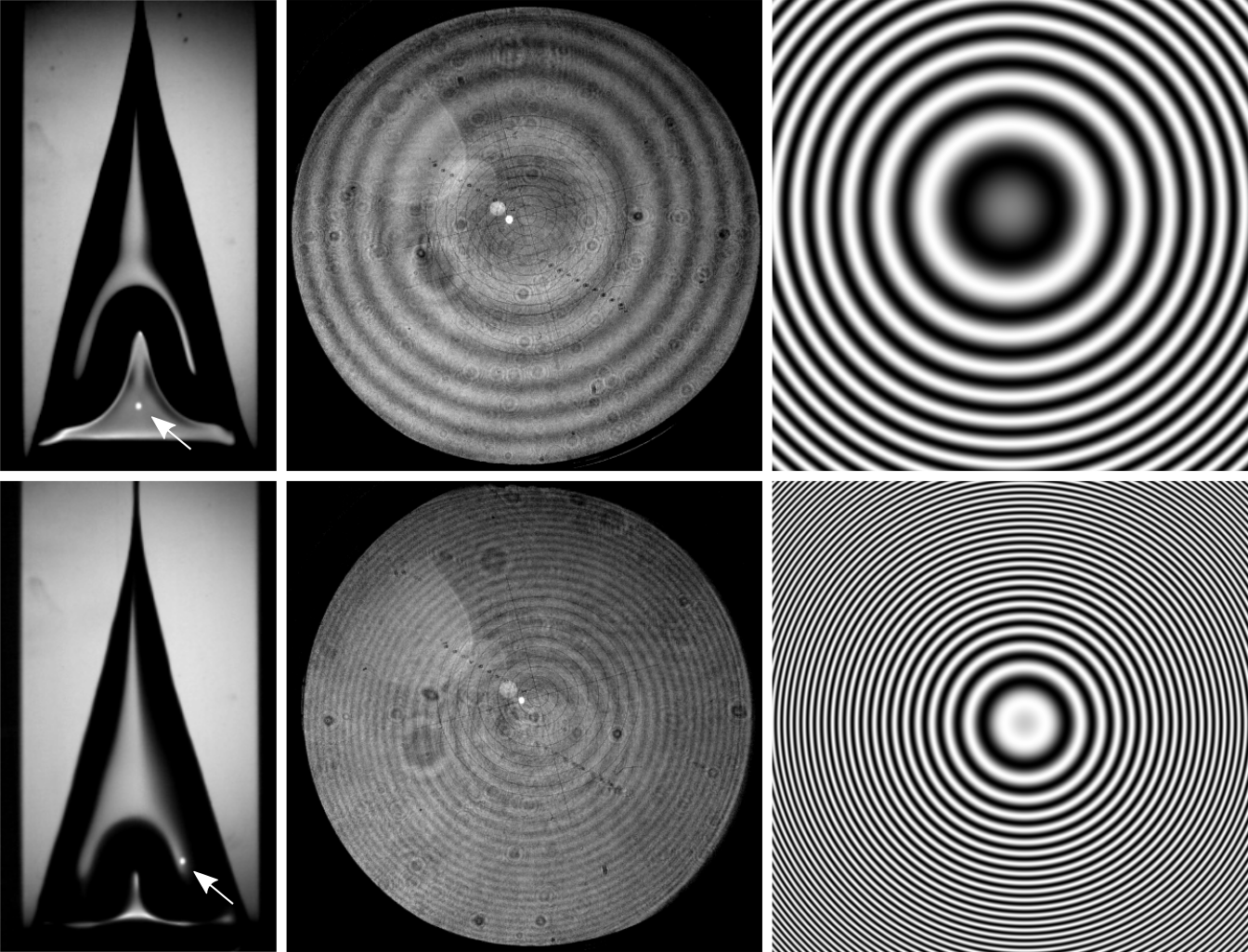}
	\caption{Thickness fitting by likelihood maximization.  An acrylic substrate travels out of a liquid of viscosity $\eta=$ 18 cP at $U=$ 17 mm/s.  Left column is an image of the entrained fluid.  The arrows pointing to focusing spots indicate location of the measurement.  Middle column: corresponding interference fringes (Haidinger's fringes; see Appendix A and Ref.~\cite{he_nagel2020optics}) captured by high-speed camera.  Right column: reconstructions of fringes using maximizing likelihood (see Appendix B), giving thicknesses of $h=$ 84.2 $\mu$m for the thin region (top row) and 285.0 $\mu$m for the thick region (bottom row).}
	\label{fig:max_likelihood}
\end{figure}

Figure~\ref{fig:max_likelihood} shows the measurements of the thin and thick regions of the dewetting layer.  The first row shows a measurement at one point of the thin, flat part.
In the image shown in the top left panel, the arrow and white spot indicate the position where the measurement is taken: near the bottom middle of the frame.  The top middle panel shows the circular fringes from the high-speed camera.  Using the method of maximum likelihood estimation allows clear identification of the interference rings; the top right panel shows the reconstructed pattern $X_0(x,y,\beta\SB{optimal})$ (see Appendix A, B and Ref.~\cite{he_nagel2020optics}).  The thickness of the entrained layer in this region is $h\textsubscript{optimal}=84.2\mu m$.  Similarly the second row shows a measurement of the thick part of the entrained fluid when it expands to touch the bottom with the measurement placed near the bottom right as shown by the arrow.  A thicker fluid layer gives rise to a much denser set of fringes, and the maximum likelihood fitting gives $h\textsubscript{optimal}=285.0\mu m$.

\begin{figure}[htp]
	\includegraphics[width=1\textwidth]{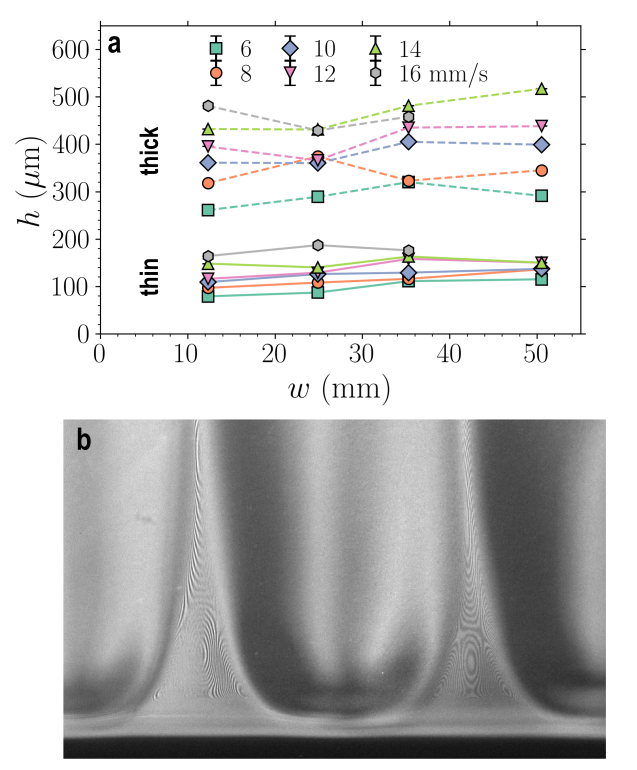}
	\caption{(a) Thickness $h$ for steady state of dewetting, versus substrate width $w$.  Markers connected by solid lines: thin part.  Markers connected by dashed lines: thick part.  Liquid viscosity $\eta=41$ cP.  (b) Zoomed-in view of the thin parts for a typical dewetting layer of water, illuminated by a sodium-vapor lamp (wavelength $\lambda=589$ nm).  Interference stripes in the thin regions indicate the flatness.}
	\label{fig:thinthickw}
\end{figure}

To see the dependence of the steady-state thickness $h$ on the substrate width $w$, I measured the thickness of both the thin and thick regions for various substrate widths.  As shown in Fig.~\ref{fig:thinthickw}a, when $w$ is varied, $h$ of the thick region fluctuates, but does not show an apparent general trend.  The thin part becomes slightly thicker ($\sim20\%$) as the width $w$ is increased by a factor of $\sim5$.  Therefore, the thickness $h$ for both the thin and thick parts are nearly independent of the plate width $w$.  In the following, I will focus on one plate width $w$ only in measuring the thicknesses.  

Figure~\ref{fig:thinthickw}b shows a typical zoomed-in image of the thin-thick alternation region, obtained using a sodium-vapor lamp (wavelength $\lambda=589$ nm).  Because of a long coherence length of the light source, interference patterns appear in the thin parts.  An order of $\sim 10$ fringes can be detected in each thin part so one can estimate the thickness variation to be $\sim3$ $\mu$m, much smaller than the thickness itself.  It shows that the thin parts are very flat, and measuring thickness at one point only is sufficient in characterizing the thin part thickness.

\begin{figure}[htp]
	\includegraphics[width=1\textwidth]{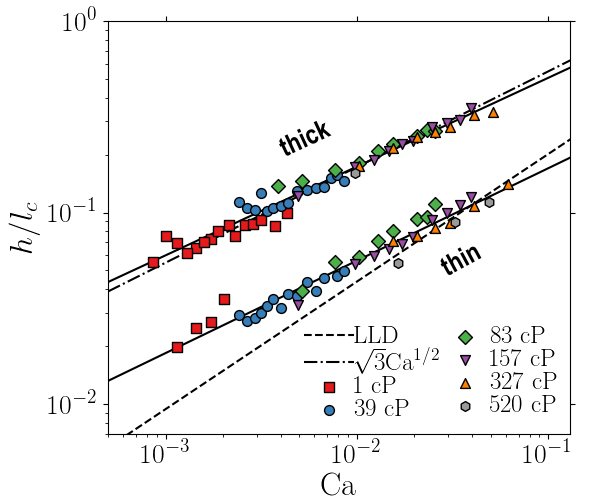}
	\caption{Thickness for dewetting layers normalized by capillary length, $h/l_c$, versus capillary number $\mathrm{Ca}$, for both the thin parts and the thick parts.  Top solid line: $h/l_c\propto \mathrm{Ca}^{0.46}$.  Bottom solid line: $h/l_c\propto \mathrm{Ca}^{0.48}$.  Substrate width $w=$ 17.1 mm}
	\label{fig:ssthinthick}
\end{figure}

For a fixed substrate width, $w$, Fig~\ref{fig:ssthinthick} shows the measurements in the dewetting steady state of the thin and thick regions as a function of liquid viscosity $\eta$ and substrate velocity $U$.  As is shown for both regions, $h$, normalized by the capillary length $l_c = \sqrt{\gamma/\rho g}$, is approximately a power law in the capillary number: $\mathrm{Ca} = \eta U/\gamma$: 
\begin{align}
	\label{eq:powerlaw}
	h/l_c \propto \mathrm{Ca}^{\alpha}.
\end{align}

For over 2 orders of magnitude in $\mathrm{Ca}$,  I find for the thick part:
\begin{align}
	\label{eq:alphathick}
	\alpha\textsubscript{thick} = 0.46 \pm 0.01.
\end{align}

This result can be understood by a similar argument as was used for the reversed situation of wetting~\cite{he_nagel2019}.  The complex-shaped liquid layer can be simplified as a wedge with an average wedge angle, $\Theta$, where it meets the substrate.  Using the result of Huh and Scriven~\cite{huh_scriven1971} that the interface velocity $U_I$ is proportional to the substrate velocity $U$,  
\begin{align}
	\label{eq:UI}
	U_I = \zeta\bigg(\frac{\eta\textsubscript{in}}{\eta\textsubscript{out}}, \Theta \bigg) U, 
\end{align}
where $\eta\textsubscript{in}$ is the viscosity of the inner fluid (water-glycerin mixture) and  $\eta\textsubscript{out}$ is the viscosity of the outer fluid (air).  In the limit of large $\eta\textsubscript{in}/\eta\textsubscript{out}$ ($10^2 \sim 10^5$) and small $\Theta$ ($<90$), it can be shown that the proportionality $\zeta$ is approximately a constant, nearly independent of $\eta\SB{in}/\eta\SB{out}$ and $\Theta$. Using the average value of the measured static receding contact angle $63^{\circ}$ as an approximation to the dynamic contact angle $\Theta$, we get $\zeta=-0.557\pm0.005$. The negative sign indicates that the flow at the interface is in the opposite direction of the substrate motion, about half in magnitude.  I argue that the thickness is selected by maximum stability of the layer, so that Eq.7 of~\cite{he_nagel2019} can be directly applied :  
\begin{align}
	\label{eq:hthick}
	h\textsubscript{thick} &= \bigg(2(1-\zeta)\frac{\eta\textsubscript{in}}{\Delta \rho g}U\bigg)^{\frac{1}{2}}\nonumber\\
	&= \sqrt{3.1}l_c \mathrm{Ca}^{\frac{1}{2}}.
\end{align}

Equation~\ref{eq:hthick} is plotted in Fig.~\ref{fig:ssthinthick} as the dashed-dotted line. As with the case of forced wetting~\cite{he_nagel2019}, the simple argument gives a reasonable fit to the data.  A rigorous derivation using lubrication theory given by Snoeijer \textit{et al.}~\cite{snoeijer2008} gives a nearly identical result as Eq.~\ref{eq:hthick}, with a pre-factor equal to $\sqrt{3}$.  Note that these two arguments both effectively applied the no-shear boundary condition at the liquid-air interface with flux conservation.  %We also emphasize that when $\eta\textsubscript{in}/\eta\textsubscript{out}$ approaches 1, i.e., when the outer fluid can no longer be neglected, the argument of maximum stability still implies a no-shear boundary condition at the interface.  Further experiments need to be carried out for cases where the outer fluid has significant contribution to the dynamics, which would test the validity of our argument in that regime.

For the thin part, for over 2 orders of magnitude in $\mathrm{Ca}$, Eq.~\ref{eq:powerlaw} also provides a good fit with 
\begin{align}
	\label{eq:thinexpo}
	\alpha\textsubscript{thin} = 0.48 \pm 0.02.
\end{align}

\begin{figure}[htp]
	\includegraphics[width=1\textwidth]{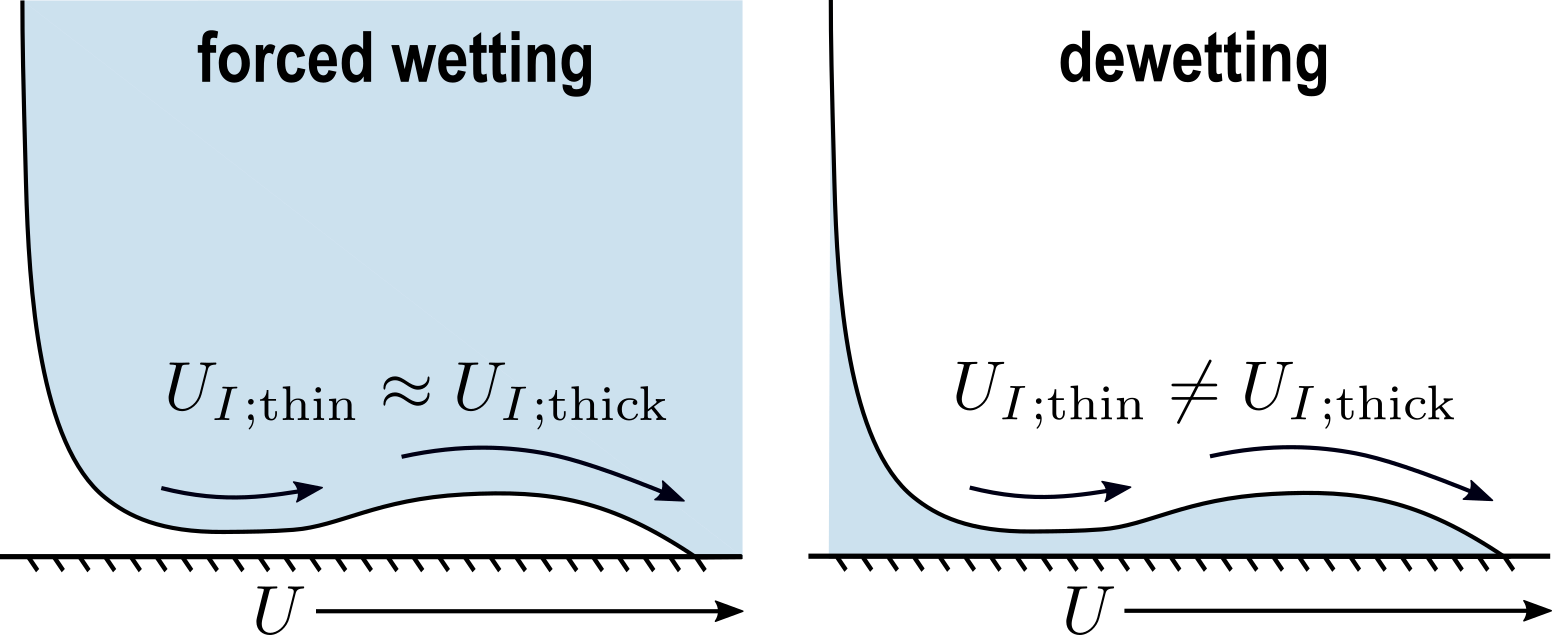}
	\caption{Left panel:  in forced wetting, the thickness variation of the entrained layer is a perturbation to the dominant bulk liquid, resulting in a roughly equal interface velocity for the thin part $U_I\textsubscript{; thin}$ and the thick part $U_I\textsubscript{; thick}$.  Right panel: in dewetting, the thickness variation of the entrained layer is expected to break the uniformity of the interface velocity, making $U_I\textsubscript{; thin} \neq U_I\textsubscript{; thick}$ in general.}
	\label{fig:U_I}
\end{figure}

Although Eq.~\ref{eq:thinexpo} is close to that for the thin parts of forced wetting~\cite{he_nagel2019}, I emphasize that the same stability argument of~\cite{he_nagel2019} does not apply since a key assumption for the argument breaks down in dewetting.  In forced wetting, we approximated the velocity of the liquid-air interface near the thin region to be equal to that of the thick region $U_I\textsubscript{; thin} \approx U_I\textsubscript{; thick}$.  As illustrated in the left panel of Fig.~\ref{fig:U_I}, this was reasonable because the thin-thick variation of the air gap is only a small perturbation to the shape of the bulk outer fluid (liquid; shaded area).  The outer fluid (liquid) is the dominant fluid except very close to the contact line~\cite{marchand2012}, hence dictating the interface velocity to be roughly uniform regardless of the gap structure.  By contrast, in the case of dewetting as illustrated in the right panel of Fig.~\ref{fig:U_I}, the inner fluid (liquid; shaded area) plays the dominant role everywhere.  The prominent thin-thick structure is expected to greatly influence the interface velocity, making the assumption of a simple, uniform interface velocity invalid.  %In fact, the interface velocity is not only not uniform, but often observed to be in the opposite directions, for the thin and thick parts in steady state dewetting.   

The thin part thickness also significantly differs from the Landau-Levich-Derjaguin theory (LLD)~\cite{landau1942, derjaguin1943, levich1962}, which has an exponent:
\begin{align}
	\label{eq:alphalld}
	\alpha\SB{LLD} = \frac{2}{3}.
\end{align}

The LLD prediction is plotted in Fig.~\ref{fig:ssthinthick} as the dashed line for comparison.  The measurements deviate from the LLD theory especially at low $\mathrm{Ca}$.  In the LLD theory, gravity and viscous dissipation are balanced, and the thickness of an \textit{infinite} liquid layer is uniquely determined by matching the meniscus shape near the bath.  Assuming in the current case a similar balance between gravity and viscous dissipation, the discrepancy suggests that the thin-part thickness $h\textsubscript{thin}$ is not selected by meniscus matching.  The existence of the contact line nearby, neighboring thicker parts and a bounding overall $V$ shape, which are not incorporated in the LLD theory, presumably play important roles.  Further modelling is required to quantitatively interpret the result Eq.~\ref{eq:thinexpo}.    

%The undulation of thickness in the horizontal direction suggests a non-uniform flow inside the liquid layer.  In fact, given the two  thicknesses, $h\textsubscript{thick}$ and $h\textsubscript{thin}$, in steady state, the relation given by conservation of flux ???WHERE DOES THIS COME FROM???:  
%\begin{align}
%	\label{fluxconserv}
%	3l_cCa &= h\textsubscript{thick}^2+ h\textsubscript{thick}h\textsubscript{thick} + h\textsubscript{thick}^2
%\end{align}
%does not hold.  This indicates that the flux is not conserved in the vertical direction.  There is siginificant transverse flow in the wetting layer. 

\begin{figure}[htp]
	\includegraphics[width=1\textwidth]{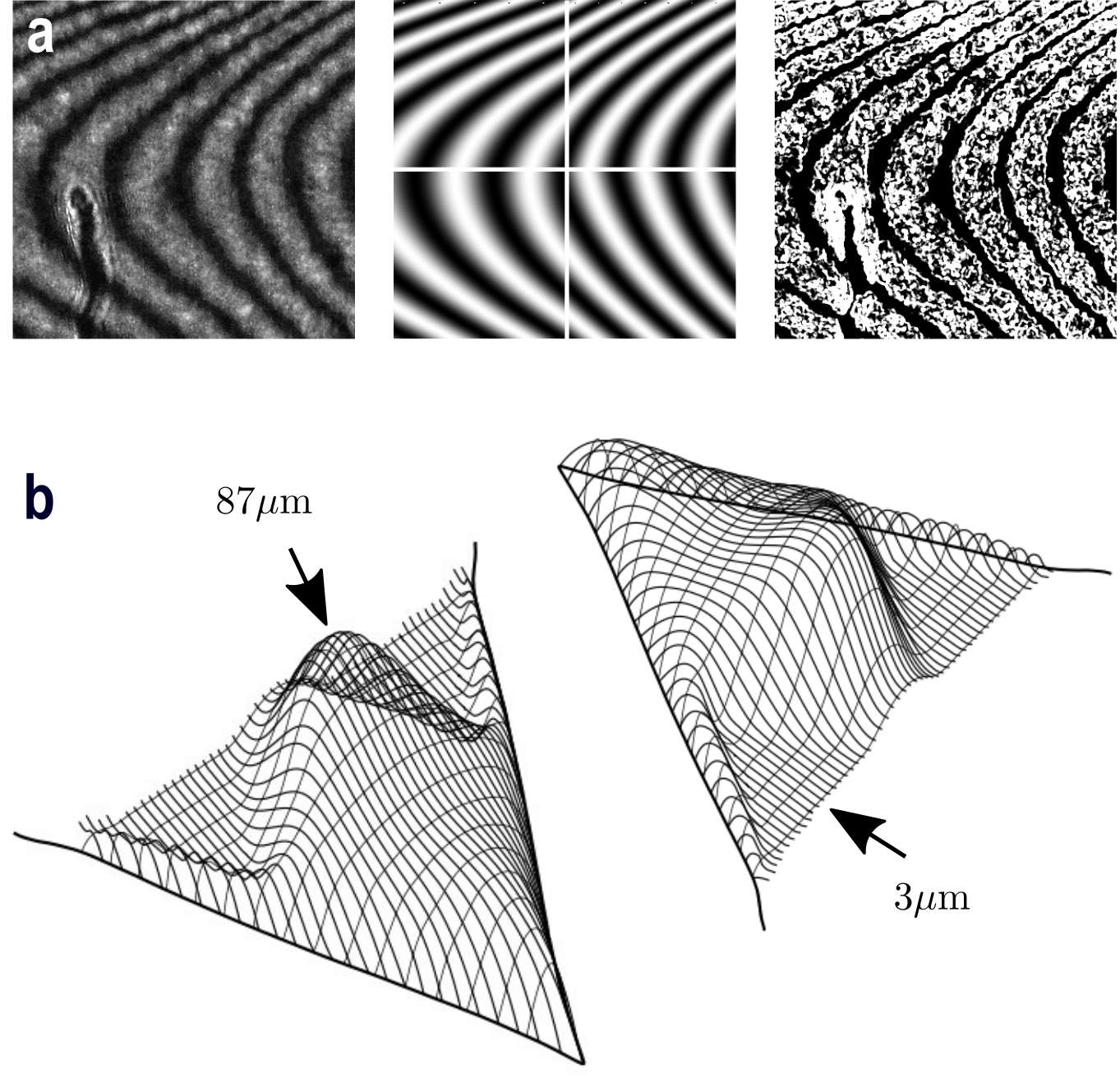}
	\caption{(a) Fringe fitting using likelihood maximization in forced wetting.  Left: interference fringes of equal height produced by part of an air film (Newton's fringes).  Middle:  Fringes reconstructed by maximum likelihood fitting.  Frame partitioned into 4 smaller parts for separate fitting.  Right: a typical local edge detection algorithm for comparison.  (b) Tomography of an air film reconstructed.  Peak thickness: 87 $\mu$m; Thin flat part: 3 $\mu$m.  Absolute thickness obtained by using multi-wavelength interference discussed in~\cite{he_nagel2019}.}
	\label{fig:max_likelihood_wetting}
\end{figure}

The method of maximum-likelihood fitting is also carried out for the case of wetting.  Figure~\ref{fig:max_likelihood_wetting}a shows a laser image of a part of the entrained air film.  There are interference fringes which are noisy and have artifacts such as a dust shadow on the bottom left.  The air film patch is divided into 4 smaller parts, each of which can be approximated by a parabolic shape.  Our fit for individual parts give satisfactory results, as shown in the middle image.  Notice the slight mismatch near the boundaries of adjacent parts.  This indicates the inadequacy of the parabolic model near the edge (rather than the inadequacy of the algorithm).  The right-most image shows a simple example of local edge detection for the same pattern, which in general cannot capture the features of main interest, and is not robust against errors.  The reconstruction of the topography of the air film is achieved by stitching these data patches together.  The result is shown in Fig.~\ref{fig:max_likelihood_wetting}b, with perspective views from two different angles.

\subsection{Onset on a wide substrate: intermediate thickness}

\begin{figure}[htp]
	\includegraphics[width=1\textwidth]{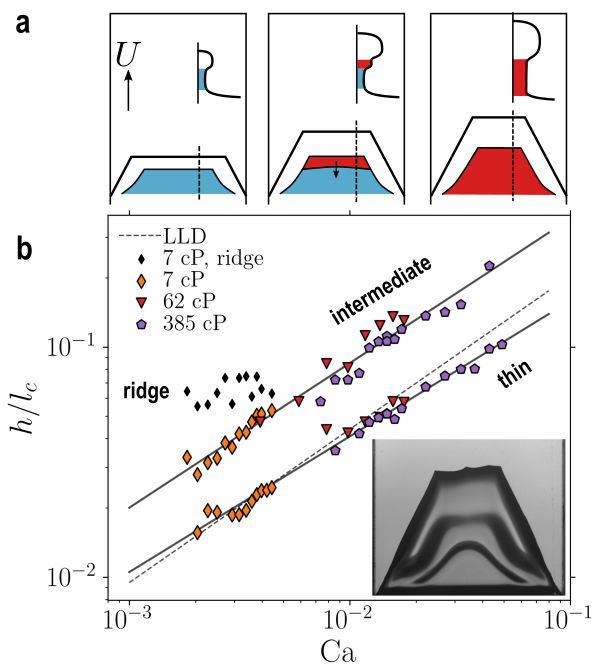}
	\caption{(a) Schematic showing the formation of a film of intermediate thickness (red) on a wide plate.  (b) Measured thickness scaled by capillary length, $h/l_c$, versus $\mathrm{Ca}$.  Top solid line:  $h/l_c\propto \mathrm{Ca}^{0.63}$.  Bottom solid line: $h/l_c\propto \mathrm{Ca}^{0.59}$.  Substrate width $w=$ 37.7 mm.  Inset:  typical image of onset film, showing the thick, intermediate, and thin regions.}
	\label{fig:thinintthickon}
\end{figure}

Previous theoretical and experimental work (e.g. ~\cite{snoeijer2006, asdf2007, asdf2008,maleki2011, gao2016}) have focused on the thickness of a non-wetting liquid film during the early stages (before completion of the triangular shape of Fig.~\ref{fig:formation}III).  They indicated that two different film thicknesses appear during the deposition: a leading ridge whose thickness is determined by the contact angle,  followed by a thin LLD film whose thickness is determined by the meniscus.  Using the interferometric method discussed above (see Appendix A and Ref.~\cite{he_nagel2020optics}), I have measured the film thickness soon after the onset, long before steady state.  The measurement was done on a wide plate with width $w=$ 37.7 mm.  I plot the result in Fig.~\ref{fig:thinintthickon}.  

During the onset stage, there is a leading thick ridge structure near the contact line, whose thickness is measured and shown as the black diamonds in Fig.~\ref{fig:thinintthickon}. The thickness of the ridge structure $h\textsubscript{ridge}$ roughly remains a constant over increasing $\mathrm{Ca}$.  This is consistent with the models of~\cite{snoeijer2006, asdf2007, gao2016}, where $h\textsubscript{ridge}$ is a function of the \textit{relative} capillary number $\mathrm{Ca}^{*}$, which does not vary much with the plate velocity (or $\mathrm{Ca}$).  I emphasize that the ridge is different from the thick part at steady state, presented in Fig~\ref{fig:ssthinthick}, in both the thickness and the dependence on $\mathrm{Ca}$.

There is an extended thin region of thickness, $h\textsubscript{thin}$, above the meniscus near the bath.  It can be fitted to $h\textsubscript{thin}/l_c\propto \mathrm{Ca}^{0.59\pm0.01}$ (bottom solid line).  This thin film is close to the LLD prediction $h\SB{LLD}/l_c = 0.946\mathrm{Ca}^{2/3}$ (dashed line), confirming previous studies of the onset of entrainment~\cite{snoeijer2008,maleki2011, gao2016}.

In addition, there appears a new region of intermediate thickness $h\textsubscript{int}$, which has not been reported in previous works.  The thin film of $h\textsubscript{thin}$ close to the LLD prediction turns out to be only transient.  It is left behind immediately after entrainment begins, lasts for a short period,  and is rapidly replaced by a region of intermediate thickness $h\textsubscript{int}$ ($h\textsubscript{thin}<h\textsubscript{int}<h\textsubscript{ridge}$). The thickness change is discontinuous.  This process is illustrated in the schematic drawing of Fig.~\ref{fig:thinintthickon}a.  The inset of Fig.~\ref{fig:thinintthickon}b shows an image of the ridge, the intermediate region and the thin region at the same time soon after entrainment.  At low plate velocities, the LLD film may never appear.

The intermediate film can be fitted to $h\textsubscript{int}/l_c \propto \mathrm{Ca}^{0.63\pm0.02}$ (top solid line).  Empirically $h\textsubscript{int}\approx2h\textsubscript{thin}$, over two decades of $\mathrm{Ca}$ range.  When the ridge thickness $h\textsubscript{ridge}$ approaches the thickness $h\textsubscript{int}$ of the intermediate region ($\mathrm{Ca} \sim 10^{-2}$ in Fig.~\ref{fig:thinintthickon}), the separation of the ridge region and the intermediate film becomes less clear.  At high $\mathrm{Ca}>10^{-2}$, the ridge structure does not appear and the layer behind the contact line assumes a monotonic thickness, which is similar to the result of~\cite{gao2016}.  (Note the intermediate region was not considered, so the monotonic film without a capillary shock occurs at $h\textsubscript{thin}=h\textsubscript{ridge}$ in their work.)  A thicker region behind the contact line will nucleate much later to form the thick parts at steady state such as those of Fig.~\ref{fig:formation}IV and Fig.~\ref{fig:thickwidth}b.

When the substrate width $w$ is small, the formation of the intermediate region during the onset is not observed (as is absent from Fig.~\ref{fig:formation}II, III for $w=$ 20.3 mm).  A thin film that can be described by the LLD theory is deposited behind a thicker ridge during entrainment before steady state.  The above observations of the intermediate region using a wider plate suggest that the plate geometry may impact the morphology of the film structure, which deserves further quantitative investigations.

\section{Conclusions}
I have presented an experimental study on various aspects of dewetting, and have systematically compared the results with those we found in forced-wetting experiments.  I have discovered a prominent structure in the layer of steady-state dewetting, consisting of well-defined thin-thick alternations transverse to the direction of substrate motion, behind a $V$-shaped contact line.  This paper draws attention to a possible instability in the spanwise direction in wetting/dewetting, which is not incorporated in most current models. 

For both wetting and dewetting,  I found quantitatively that the normal relative velocity is larger during the onset than it is at steady state, which extends the previous observations and is different from a fixed maximum contact-line speed in other wetting/dewetting geometries. 

To characterize and  quantify precisely the thin-thick structure in the dewetting layer, I developed a method, combining interference information from varying the angle of incidence and pattern fitting with maximum likelihood estimation.  Power-law relationships are found between layer thickness $h$ and capillary number $\mathrm{Ca}$ over two decades of $\mathrm{Ca}$ range, for different parts of steady state.  The thickness of steady state thin part in dewetting differs from various existing models.  The new pattern-fitting algorithm also helps to reconstruct the topography of the air layer in forced wetting.  

Lastly, onset of dewetting entrainment has been examined and I found a new region, whose thickness is in between two known regions predicted and observed in various previous studies.  

This work shows that dynamic partial wetting is far more complex than accounted for in various simple models.  Future work is needed to quantify and understand the contact-line velocity variation as well as the mechanism for thickness selection in both the onset and steady state.  Further experiments on wetting in two-liquid systems, where both liquids contribute significantly, can help to examine and clarify the argument of stability.      

\section{Acknowledgements}
I am deeply indebted to Sidney R. Nagel for his advising and mentoring.  I thank Anthony LaTorre for extensive discussions on various computational techniques on likelihood maximization.  I also thank Amy Schulz for kindly coordinating a financial support. 

I am particularly grateful to Chloe W. Lindeman for taking the time to capture the side-view images of sample drops which made Fig.~\ref{fig:gamma_CA}b possible, as well as Nidhi Pashine for transferring key backup data files, while I did not have access to the lab and the data.  

The work was primarily supported by the University of Chicago MRSEC, funded by the National Science Foundation under Grant No. DMR-1420709 and NIST, Center for Hierarchical Materials Design (CHiMaD) (70NANB14H012).

\section{Appendix A. Measurement of absolute thickness: principle}

\begin{figure}[htp]
	\includegraphics[width=0.5\textwidth]{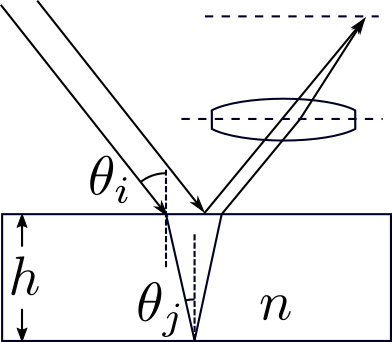}
	\caption{Schematic of interference produced by a beam with angle of incidence $\theta_i$ onto a sample of thickness $h$.  Light reflected from top surface at $\theta_i$ and bottom surface at $\theta_j$ interfere at the focal plane of a lens placed above.}
	\label{fig:principle}
\end{figure}

When a parallel beam of light is incident upon a transparent sample of thickness $h$ with an angle of incidence $\theta_i$, the reflected beam from the top surface at angle $\theta_i$ and the one from the bottom surface at $\theta_j$ are brought together by a lens placed above, as shown in Fig.~\ref{fig:principle}.  Interference occurs at the focal plane of the lens.  Considering the phase change upon reflection, the optical path difference is $2nh\cos\theta_ j+\lambda/2$, where $n$ is the refractive index of the sample and $\lambda$ is the wavelength of the light source.  The intensity of interference depends on the angle of incidence $\theta_i$.  When the interference is completely destructive:
\begin{align}
	\label{eq:opd}
	2nh\cos \theta_j = m\lambda,
\end{align}
where $m$ is an integer indicating the order of the destructive interference.  All the rays that produce a single dark fringe correspond to the same value of $\theta_j$.  These fringes are fringes of equal inclination.  
When the angle of incidence is changed from $\theta_{i1}$ to $\theta_{i2}$ such that the order of interference increases by $\Delta m =1$, from Eq.~\ref{eq:opd} the corresponding $\theta_{j1}$ and $\theta_{j2}$ satisfy:
\begin{align}
	2nh(\cos \theta_{j2}-\cos \theta_{j1}) = \lambda\nonumber.
\end{align}

Since $\theta_{j}$ is related to $\theta_{i}$ through Snell's law: 
\begin{align}
	\label{eq:snell}
	\sin \theta_i &= n\sin \theta_j,
\end{align}
we have
\begin{align}
	\label{eq:i2h}
	2nh(\cos\arcsin\frac{\sin \theta_{i2}}{n} -\cos\arcsin\frac{\sin \theta_{i1}}{n}) = \lambda.
\end{align}

If $\theta_{i1}$ and $\theta_{i2}$ can be measured, Eq.~\ref{eq:i2h} gives a determination of $h$ if the index of refraction, $n$, is known.  Notice that no small-angle approximation has been assumed, making the above analysis valid for arbitrary angles of incidence.

In particular, in our setup described in Ref.~\cite{he_nagel2020optics}, the interference pattern is a set of concentric rings at the focal plane.  $\theta_j$ is given by:
\begin{align}
	\label{eq:x2theta}
	\sin\theta_j &= \frac{1}{n}\sqrt{\frac{(x-x_c)^2+(y-y_c)^2}{(x-x_c)^2+(y-y_c)^2+f^2}},
\end{align}
where $f$ is the focal length of the convex lens, $(x, y)$ is the coordinate at the focal plane and  $(x_c, y_c)$ is the center of the rings.

\section{Appendix B. Data fitting: likelihood maximization}
In the thickness measurements, one needs to convert data images of interference fringes to thickness information.  It can be extremely difficult to extract fringe patterns from noisy data images.  Local edge detection algorithms often perform poorly for patterns whose length scales span multiple orders of magnitude in the presence of a wide range of noise and artifacts (\textit{e.g.}, shadows, lens flares, etc.).  An example is shown in the third frame of Fig.~\ref{fig:max_likelihood_wetting}a.

Since the physical model, \textit{i.e.}, the relation between fringe configuration and thickness $h$, is known, I approach this problem using likelihood maximization.  For a data image the measured pixel intensity of coordinate $(x, y)$ is $X(x, y)$.  The parameters of the model are denoted as $\beta$, the log-probability $\log P$ for all pixels of the data image taking the current values is: 
\begin{align}
	\label{eq:ll}
	\log P &= \log \prod_{x,y}P(X(x,y),\beta) \nonumber\\
	&= \log \prod_{x,y}P(X(x,y)|\beta)P(\beta)\nonumber\\
	&\propto \sum_{x,y}\log P(X(x,y)|\beta).
\end{align}
In the last step above, $P(\beta)$ is omitted since the model parameter vector $\beta$ is not a random vector.  The best $\beta$ that fits the data is that which maximizes the log-likelihood $l(\beta) \equiv \sum_{x,y}\log P(X(x,y)|\beta)$ (viewed as a function of $\beta$): 
\begin{align}
	\label{eq:beta}
	\beta\textsubscript{optimal} &= \argmax_{\beta}l(\beta).
\end{align}

The expression of $l(\beta)$ depends on how the pixel fluctuation is modelled.  Consider the simple case of normal distribution $P(X(x,y)|\beta) \propto \exp[-(X(x,y)-X_0(x,y,\beta))^2/\sigma^2]$ where $X_0(x,y,\beta)$ is the expected pixel intensity from the physical model given a particular vector $\beta$.  Then we have:
\begin{align}
	\label{eq:gaussian}
	l(\beta) &\propto \sum_{x,y}-(X(x,y)-X_0(x,y,\beta))^2.
\end{align}
Thus, from Eq.~\ref{eq:beta} 
\begin{align}
	\label{eq:leastsq}
	\beta\textsubscript{optimal} &= \argmin_{\beta}\sum_{x,y}(X(x,y)-X_0(x,y,\beta))^2.
\end{align}
Therefore, under the assumption of normal distribution of pixel intensity, finding the optimal parameter $\beta$ amounts to a least-square regression.

In the case of dewetting, we take $h$, $x_c$ and $y_c$ as 3 fitting parameters.  Combining Eq.~\ref{eq:opd}, Eq.~\ref{eq:snell} and Eq.~\ref{eq:x2theta}, the expected intensity $X_0(x,y,\beta))$ is given by: 

\begin{empheq}[left = \empheqlbrace]{align}
	\label{eq:model_dewetting}
	\begin{split}
		X_0(x,y,\beta)) &= \frac{1}{2}+\frac{1}{2}\cos(\frac{2\pi}{\lambda}2nh\cos\theta_j+\pi)\\
		\sin\theta_j &= \frac{1}{n}\sqrt{\frac{(x-x_c)^2+(y-y_c)^2}{(x-x_c)^2+(y-y_c)^2+f^2}}\\
		\beta &= (h, x_c, y_c).
	\end{split}
\end{empheq}

Substituting Eq.~\ref{eq:model_dewetting} into Eq.~\ref{eq:leastsq} gives the expression of $\beta\textsubscript{optimal}=(h\textsubscript{optimal},x_c\SB{,optimal}, y_c\SB{,optimal})$.  Since the right-hand-side of Eq.~\ref{eq:leastsq} is highly non-convex, $\beta\textsubscript{optimal}$ is found by brute-force searching through all nodes in $(h, x_c, y_c)$ parameter space, with step resolution $\delta h = \lambda/(4n)$, $\delta x_c = \delta y_c = 1$ pixel.  With known centers $(x_{c,a}, y_{c,a})$ and $(x_{c,b}, y_{c,b})$, an exhaustive search in the parameter space $\beta = (h, h+p\lambda/2n, n)$ to maximize the joint likelihood of the two frames (summation over all pixels for two frames in Eq.~\ref{eq:leastsq}) gives the optimal $n$.

Similarly for the case of forced wetting, I use normal incidence only and model the interference fringes of equal height.  The patterned area is divided into smaller parts, whose thickness can be approximated by a quadratic expansion.  This is shown in Eq.~\ref{eq:model_wetting}.  Since there are 6 components to optimize in $\beta$ of this model,  I use a basin-hopping minimizing algorithm instead of brute-force searching.
\begin{empheq}[left = \empheqlbrace]{align}
	\label{eq:model_wetting}
	\begin{split}
		X_0(x,y,\beta)) &= \frac{1}{2}+\frac{1}{2}\cos(\frac{2\pi}{\lambda}2nh(x,y,\beta)+\pi)\\
		h(x,y,\beta) &=\beta_1x^2 +\beta_2 y^2+\beta_3xy\\
		&+ \beta_4x+\beta_5y\\
		&+ \beta_6\\
		\beta &= (\beta_1,\beta_2,\beta_3,\beta_4,\beta_5,\beta_6).
	\end{split}
\end{empheq}

\bibliography{refs}

\end{document}